\documentclass{elsart}

\usepackage{epsfig}

\usepackage{amssymb}

\begin{document}

\def\FA{{\em FeynArts }}
\def\FC{{\em FormCalc }}
\def\LT{{\em LoopTools }}
\def\FF{{\em FF }}
\def\MA{{\em Mathematica }}
\def\FM{{\em Form }}
\def\FN{{\em Fortran }}

\def\beq{\begin{equation}}
\def\eeq{\end{equation}}
\def\bea{\begin{eqnarray}}
\def\eea{\end{eqnarray}}

\def\d  {{\rm d}}
\def\M  {{\mathcal M}}

\begin{frontmatter}



\title{Calculating two- and three-body decays with \FA and \FC}


\author{Michael Klasen}

\address{II.\ Institut f\"ur Theoretische Physik, Universit\"at Hamburg,
         Luruper Chaussee 149, D-22761 Hamburg, Germany}

\begin{abstract}
The Feynman diagram generator \FA and the computer algebra program \FC allow
for an automatic computation of $2\to 2$ and $2\to 3$ scattering processes
in High Energy Physics. We have extended this package by four new kinematical
routines and adapted one existing routine in order to accomodate also two- and
three-body decays of massive particles. This makes it possible to compute
automatically two- and three-body particle decay widths and decay energy
distributions as well as resonant particle production within the Standard Model
and the Minimal Supersymmetric Standard Model at the tree- and loop-level. The
use of the program is illustrated with three standard examples:
$h\to b\bar{b}$, $\mu\to e\bar{\nu}_e\nu_\mu$, and $Z\to\nu_e\bar{\nu}_e$.
\end{abstract}

\begin{keyword}
Feynman diagrams \sep perturbative calculations \sep decays

\PACS 02.70.-c \sep 13.35.-r \sep 13.38.-b \sep 14.80.-j
\end{keyword}
\end{frontmatter}

\vspace*{-16cm}
DESY 02-179 \\
hep-ph/0210426
\vspace*{14cm}


\section{Introduction}
\label{sec:1}

The analytical calculation of Feynman diagrams and their subsequent numerical
evaluation represent the standard technique for calculating cross sections
of scattering processes and widths of particle decays in High Energy Physics.
Unfortunately, the number of diagrams to be computed grows rapidly with the
number of external particle legs, internal particle loops, and exchanged
particles. This makes an automatic generation and computation of Feynman
diagrams a highly desirable feature, not only to reduce the time needed for
the calculations, but also to eliminate possible sources of errors.

For more than ten years, a well-tested and easy-to-use Feynman diagram
generator has existed in the form of the \MA package \FA \cite{Kublbeck:xc}.
More recently, a fast and equally reliable computer algebra program for the
evaluation of the Dirac and color traces inherent in the generated diagrams
has been implemented within \FM (\FC$\!\!$) \cite{Hahn:1998yk}. Together with
\LT \cite{Hahn:1998yk} (an adapted version of the \FN program \FF
\cite{vanOldenborgh:1989wn} for the numerical evaluation of tensor loop
integrals), the \FA and \FC program package allows for an automatic
computation of tree- and loop-level $2\to2$ and $2\to3$ scattering processes
within the Standard Model and the Minimal Supersymmetric Standard Model. 
Since the phase space for each particular scattering process depends on the
number of in- and out-going external particles, it is best evaluated directly
within \FN with specific kinematical routines.
So far, only $2\to 2$ and $2\to 3$ scattering processes have been implemented
within \FA and \FC$\!\!$. The aim of this Paper to extend their applicability
also to two- and three-body decays of massive particles and to resonant
particle production. We provide four new kinematical routines and one adapted
routine, which allow for the computation of the total two- and three-body
decay widths and of the energy distribution of one decay product for
three-body decays.

The remainder of this Paper is organized as follows: In Sec.\ \ref{sec:2} we
collect the well-known kinematical formul\ae \ for two- and three-body decays.
Their implementation in the new and modified \FN routines is described in
Sec.\ \ref{sec:3}. As a test of the new routines, we calculate in Sec.\
\ref{sec:4} three standard examples for scalar, fermion, and vector boson
decays. A short summary is given in Sec.\ \ref{sec:5}.

\section{Kinematics for two- and three-body decays}
\label{sec:2}

The decay width of a heavy particle with mass $m_1$,
\beq
 \d\Gamma_i={1\over F}\,|\M|^2\,\d\Phi_i,
\eeq
depends on the flux factor $F$, {\it i.e.} the number of decaying particles per
unit volume, the squared invariant scattering amplitude $|\M|^2$, and the
phase space volume $\d\Phi_i$ for $i$ final-state particles. In the rest frame
of the decaying particle, $k_1=(m_1,0,0,0)$, the flux factor is simply
\beq
 F=2m_1.
\eeq
The four-momenta of two decay products with masses $m_2,\,m_3$,
\bea
 k_2&=&(E_2,~~\,|k_2|\sin\theta,\,0,~~\,|k_2|\cos\theta), \nonumber \\
 k_3&=&(E_3,-   |k_3|\sin\theta,\,0,-   |k_3|\cos\theta),
\eea
are uniquely defined by their energies $E_i^2=|k_i|^2+m_i^2$, squared
three-momenta
\beq
 |k_2|^2=|k_3|^2={(m_1^2-m_2^2+m_3^2)^2\over4m_1^2}-m_3^2,
\eeq
and by the center-of-mass scattering angle $\theta$. The phase space volume for
two-body decays is then \cite{byckling}
\bea
 \d\Phi_2 &=& {\d^3k_2\over(2\pi)^32E_2}{\d^3k_3\over(2\pi)^32E_3}
 (2\pi)^4\delta^{(4)}(k_1-k_2-k_3) \nonumber \\
 &=& {1\over(2\pi)^2}\,{|k_2|\over4m_1}\,\d\cos\theta\,\d\phi,
\eea
where the dependence on the azimuthal angle $\phi$ integrates to $2\pi$. The
total two-body decay width is obtained by integrating over the polar angle
$\theta$. [Note that in the rest frame of the decaying particle, the
distribution in the polar angle $\theta$ depends on the arbitrary definition
of a reference axis.]

For three-particle decays, the phase space volume is \cite{byckling}
\bea
 \d\Phi_3 &=& {\d^3k_2\over(2\pi)^32E_2}{\d^3k_3\over(2\pi)^32E_3}
 {\d^3k_4\over(2\pi)^32E_4}(2\pi)^4\delta^{(4)}(k_1-k_2-k_3-k_4) \nonumber \\
 &=& {1\over(2\pi)^4}\,{1\over8}\,\d E_2\,\d\phi_2\,\d E_4\,\d\cos\theta_4.
\eea
Here, the $\delta$-function has been used to eliminate the integration over
$k_3$ and $\theta_2$, and the trivial integration over the azimuthal
angle $\phi_4$ has already been performed. For three-body decays, we will not
only consider total decay widths, but also distributions in the decay energy
$E_4$.

While the standard version of \FA and \FC diverges for $2\to 2$ and $2\to 3$
cross sections, which proceed through an intermediate $s$-channel resonance,
our modified version makes it possible to calculate also these resonance cross
sections using the formula
\beq
 \sigma_{ab\to R\to X}(\sqrt{s})={4\,\pi\,\Gamma(R\to ab)\,\Gamma(R\to X)\over
 (s-m_R^2)^2+m_R^2\,(\Gamma_{\rm tot}^R)^2},
\eeq
where $s$ is the squared center-of-mass energy of the initial state particles
$a$ and $b$, $m_R$ is the mass of the resonance $R$, and the decay widths are
calculated as described above.

\section{Description of the new and modified \FN routines}
\label{sec:3}

The phase space integrals described in the previous Section have been
implemented in \FA and \FC with two new \FN routines for two- and three-body
decays,
\begin{itemize}
 \item {\tt 1to2.F},
 \item {\tt 1to3.F},
\end{itemize}
and two corresponding common blocks,
\begin{itemize}
 \item {\tt 1to2.h},
 \item {\tt 1to3.h}.
\end{itemize}
The new \FN routines have been adapted from the existing routines for
$2\to2$ and $2\to3$ particle scattering by modifying variable and
center-of-mass energy definitions, the flux factor, and the initial-state
four-momenta. The \FN routine
\begin{itemize}
 \item {\tt num.F},
\end{itemize}
which handles the averaging and the summation over initial and final state
spins, has been modified to accomodate single-particle initial states. For this
purpose, the routine {\tt num.F} checks if the new preprocessor variable
{\tt DECAY} has been defined in {\tt 1to2.F} or {\tt 1to3.F} or not.%
\footnote{The new \FN routines can be obtained from klasen@mail.desy.de.}

\section{Three examples: $h\to b\bar{b}$, $\mu\to e\bar{\nu}_e\nu_\mu$, and $Z\to\nu_e\bar{\nu}_e$}
\label{sec:4}

In order to demonstrate that the new kinematical routines work properly within
the extended package of \FA and \FC and are straightforward to use, we compute
in this Section three standard decay widths. Note that the \FC routine
{\tt process.h} has to be adapted to every considered scattering process.
In particular, the correct phase space generator has to be included at the
end of {\tt process.h}. However, the calling structure of the main program
{\tt run.F} has not been changed. For more details we refer the reader to the
\FA \cite{Kublbeck:xc} and \FC \cite{Hahn:1998yk} manuals.

\subsection{$h\to b\bar{b}$}
The isotropic, scalar two-body decay $h\to b\bar{b}$ represents the dominant
Higgs decay mode for a Standard Model Higgs boson with mass below $m_H=140$
GeV. Using the default parameter settings of \FA and \FC$\!\!$ and the Standard
Model initialization file {\tt sm\_ini.F}, we obtain the decay widths listed
in Table \ref{tab:1}.
\begin{table}
\begin{tabular}{|c|c|}
\hline
$m_H/$GeV & $\Gamma(H\to b\bar{b})/$MeV \\
\hline
\hline
100 & 4.15 \\
110 & 4.58 \\
120 & 5.00 \\
130 & 5.42 \\
140 & 5.85 \\
150 & 6.27 \\
\hline
\end{tabular}
\caption{\label{tab:1}Decay widths of a Standard Model Higgs boson into a pair
 of bottom quarks for six different Higgs masses $m_H$.}
\end{table}
Identical results are obtained with the tree-level formula \cite{peskin}
\beq
 \Gamma(H\to b\bar{b}) = {\alpha\,N_C\,m_H\,m_b^2\over8\,\sin^2\theta_W\,m_W^2} \, \left(1-{4m_b^2\over m_H^2}\right)^{3/2}
\eeq
where $1/\alpha=137$ is the inverse of the electromagnetic fine structure
constant, $\sin^2\theta_W=1-m_W^2/m_Z^2$ is the electroweak mixing
angle, $N_C=3$ accounts for the color degree of freedom, $m_H$, $m_Z=91.1882$
GeV, and $m_W=80.419$ GeV are the masses of the Higgs, $Z$-, and $W$-bosons,
and $m_b=4.7$ GeV is the mass of the bottom quark. We have checked that the
cross section differential in the scattering angle $\theta$ is indeed constant.

\subsection{$\mu\to e\bar{\nu}_e\nu_\mu$}

\begin{figure}
 \epsfig{file=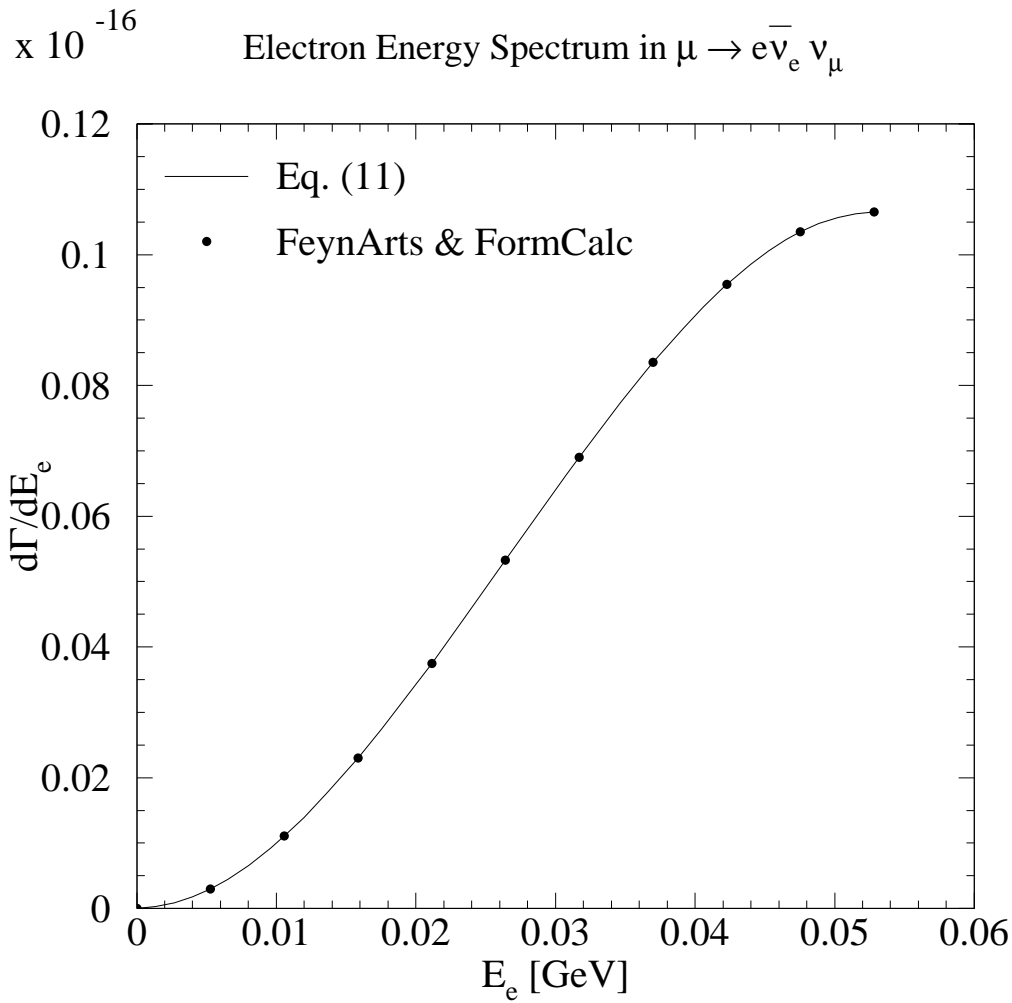,width=\textwidth}
 \caption{\label{fig:1}Electron energy spectrum of the decay width for
 muons decaying into electrons and neutrinos.}
\end{figure}

In the Standard Model, the three-body decay of muons with mass $m_\mu=105.658$
MeV into electrons and neutrinos proceeds through a virtual $W$-boson and
accounts for almost 100\% of the muonic decay width. With the default
parameter settings of \FA and \FC$\!\!$, we obtain
\beq
 \Gamma(\mu\to e\bar{\nu}_e\nu_\mu)=2.8\times10^{-10} \ {\rm eV}
\eeq
which agrees with the result obtained from the lowest order formula
\cite{halzen}
\beq
 \Gamma(\mu\to e\bar{\nu}_e\nu_\mu)={G^2\,m_\mu^5\over192\,\pi^3},
\eeq
where we have averaged over the initial-state spin, Fermi's constant $G$ has
been computed from $G=\pi\,\alpha/(\sqrt{2}\-\,\sin\theta_W^2\,m_W^2)$, and
the electron mass has been neglected.
Furthermore, the calculated muon life time $\tau=\hbar/\Gamma=2.3\times10^{-6}$
s is in good agreement with the measured value of $2.2\times10^{-6}$ s. As
mentioned above, for three-body decays we consider also the energy
distribution of one of the decay products, in this case of the observed
electron \cite{halzen}
\beq
 {\d\Gamma\over\d E_e}(\mu\to e\bar{\nu}_e\nu_\mu)=
 {G^2\,m_\mu^2\over12\,\pi^3}\,E_e^2\,\left(3-{4\,E_e\over m_\mu}\right).
 \label{eq:1}
\eeq
As can be seen in Fig.\ \ref{fig:1}, the numerical evaluation of Eq.\
(\ref{eq:1}) agrees very well with the result of \FA and \FC $\!\!$.

\subsection{$Z\to\nu_e\bar{\nu}_e$}

In our final example, we consider the invisible branching ratio of the
$Z$-boson in the Standard Model, which represents an important boundary
condition for the number of neutrinos realized in nature. The tree-level
formula \cite{halzen}
\beq
 \Gamma(Z\to\nu_e\bar{\nu}_e)=
 {\alpha\,m_Z\over24\,\sin^2\theta_W\,\cos^2\theta_W}
\eeq
and our modified program package \FA and \FC unanimously lead to the numerical
result
\beq
 \Gamma(Z\to\nu_e\bar{\nu}_e)=160 \ {\rm MeV}.
\eeq
For three generations of neutrinos and employing the measured total width of
the $Z$-boson, $\Gamma_{\rm tot}=2.49$ GeV, this translates into an invisible
branching ratio of 
\beq
 {\rm BR}_{\rm inv} = 3 {\Gamma(Z\to\nu_e\bar{\nu}_e)\over\Gamma_{\rm tot}}
 = 19.3\%,
\eeq
which is close to the experimental value of 20.0\%.

\section{Summary}
\label{sec:5}

In this Paper, we have described the implementation of two- and three-body
decays and resonance cross sections in the computer algebra package \FA and
\FC$\!\!$. We have provided
four new \FN routines, which generate the phase space for two- and
three-particle decays, and we have modified one existing routine in order to
account for correct spin averages of one-particle initial states. While the
new routines have been obtained in a straight-forward way and are easy to use,
they will hopefully serve a large variety of purposes and a wide community in
High Energy Physics to obtain interesting new results for tree- and loop-level
particle decays or resonance cross sections.

\section*{Acknowledgments}

This work has been supported by the Deutsche Forschungsgemeinschaft through
Grant No.\ KL~1266/1-3.



\end{document}